\documentclass[12pt]{article}
\usepackage{scicite}
\usepackage{times}
\usepackage{graphicx}
\usepackage{color}

\newenvironment{sciabstract}{%
\begin{quote} \bf}
{\end{quote}}

\newcounter{lastnote}
\newenvironment{scilastnote}{%
\setcounter{lastnote}{\value{enumiv}}%
\addtocounter{lastnote}{+1}%
\begin{list}%
{\arabic{lastnote}.}
{\setlength{\leftmargin}{.22in}}
{\setlength{\labelsep}{.5em}}}
{\end{list}}

\title{Experimental Proof of a Magnetic Coulomb Phase}

\author{Tom Fennell$^{1\ast}$, 
P. P. Deen$^{1}$,
A. R. Wildes$^{1}$,
K. Schmalzl$^{2}$,\\
D. Prabhakaran$^{3}$,
A. T. Boothroyd$^{3}$,\\
R. J. Aldus$^{4}$,\\
D. F. McMorrow$^{4}$,
S. T. Bramwell$^{4}$
\\
\normalsize{$^{1}$Institut Laue-Langevin, Grenoble, France}\\
\normalsize{$^{2}$IFF, Forschungszentrum Juelich GmbH,
JCNS at ILL, Grenoble, France}\\
\normalsize{$^{3}$Clarendon Laboratory, Oxford Physics, Park Road, Oxford, UK}\\
\normalsize{$^{4}$London Centre for Nanotechnology and Department}\\ 
\normalsize{of Physics and Astronomy, University College London, UK}\\
\\
\normalsize{$^\ast$To whom correspondence should be addressed; E-mail: fennell@ill.fr}
}
\date{}

\begin{document} 


\baselineskip24pt


\maketitle


\begin{sciabstract}
Spin ice materials are magnetic substances in which the spin directions map onto hydrogen positions in water ice. Recently this analogy has been elevated to an electromagnetic equivalence, indicating that the spin ice state is a Coulomb phase, with magnetic monopole excitations analogous to ice's mobile ionic defects. No Coulomb phase has yet been proved in a real magnetic material, as the key experimental signature is difficult to resolve in most systems. Here we measure the scattering of polarised neutrons from the prototypical spin ice Ho$_2$Ti$_2$O$_7$. This enables us to separate different contributions to the magnetic correlations to clearly demonstrate the existence of an almost perfect Coulomb phase in this material. The temperature dependence of the scattering is consistent with the existence of deconfined magnetic monopoles connected by Dirac strings of divergent length.  

\end{sciabstract}

\newpage

Cooperative paramagnets are of topical interest on account of their diverse phases~\cite{leenature,prl1,ramirez,science}, phase transitions~\cite{pickles} and excitations~\cite{u1,jason, monopoles,ludovic}, as well as serving as models for micromagnetic arrays~\cite{schiffer}, chemical structures~\cite{boron} and colloidal films~\cite{colloids}. Recent theoretical work~\cite{clh,isakov} has recognised that the cooperative paramagnetic state should differ fundamentally from the conventional paramagnetic state in the form of the spin correlations at large distance. In a conventional paramagnet the spin-spin correlation function decays like a screened Coulomb interaction $\frac{1}{r}e^{-\kappa r}$ where $\kappa$ is the inverse correlation length, while in a large class of cooperative paramagnets, it is predicted to decay like a dipolar interaction $\sim \nabla_{\bf r} \nabla_{{\bf r}'} \frac{1}{{\bf r} - {\bf r}'}$. This behaviour is a consequence of  `ice-rule' type constraints on the local spin configurations as was previously noted for the analogous case of polarisation correlations in ice-rule paraelectrics~\cite{yam}.  The pseudo-dipolar correlations in direct (`real') space Fourier transform into a set of so called `pinch point' singularities in reciprocal space. These features, which resemble bow ties, afford an unambiguous signature that may be characterized in a scattering experiment. However, although pseudo-dipolar correlations have been {\it predicted} for spin ices~\cite{clh,isakov}, frustrated antiferromagnets~\cite{zhz,clh,isakov} and dimer systems~\cite{alet,huse,normand}, pinch points have only been clearly {\it observed} in the field-induced kagome ice phase of spin ice~\cite{meenp,tabata}, their absence in other experimental systems being notable~\cite{ballou,leenature,mgcr}.    

In the particular case of spin ices such as Ho$_2$Ti$_2$O$_7$~\cite{prl1,science}, unpolarised neutron scattering has established that the dipolar spin ice model, in which the rare earth ions are coupled by the dipolar interaction and competing superexchange, gives a very accurate description of bulk and microscopic properties~\cite{newprl,tarasprl}. However, a projective equivalence between the long ranged dipolar interaction and a near-neighbour ferromagnetic interaction, in which the ice rules alone constrain the ground state spin configurations~\cite{isakov,iceisice}, implies that the spin-spin correlations of Ho$_2$Ti$_2$O$_7$ will be very close to those of the full ice rule manifold of the near neighbour spin ice model~\cite{prl1}.  Despite the strong expectation of the existence of pinch points in the zero field spin ice state~\cite{clh,isakov}, none have yet been resolved~\cite{newprl,meeprb,meenp,clancy,tarasprl}.

It is particularly important to establish the existence of pseudo-dipolar correlations in a real spin ice as Castelnovo, Moessner and Sondhi~\cite{monopoles} have predicted that thermal defects in the spin ice state are equivalent to ionic defects in water ice, not only in a statistical sense (which follows from the spin ice mapping~\cite{prl1}), but also at the level of  {\it electrodynamics}. In the spin ice ground states the magnetization ${\bf M}({\bf r})$ and the magnetic ${\bf H}$-field are divergence free, but thermal excitations create deconfined sources and sinks in ${\bf M}$ and ${\bf H}$ that may be described as magnetic `monopoles'~\cite{monopoles,ludovic}. The divergence free constraint on the coarse grained ${\bf M}({\bf r})$ is a consequence of the pseudo-dipolar spin {\it correlations}, while the Coulombic attraction of  the excitations is a consequence of physical dipolar {\it interactions} between spins. As dipolar interactions are not in doubt, the observation of dipolar correlations (pinch points) would therefore prove the existence of the postulated `magnetic Coulomb phase'~\footnote{In this context we take a Coulomb phase to be a thermodynamic phase in which the elementary excitations are monopoles.}. 

Neutron scattering, when appropriately energy integrated, estimates the scattering function 
$S^{\alpha\beta}({\bf Q})$ in reciprocal space (here $\alpha,\beta = x,y,z$), which is the required Fourier transform of the thermally averaged two-spin correlation function. Our polarised neutron experiments were configured to measure two independent components of the tensor $S^{\alpha\beta}({\bf Q})$ that we label as spin flip (SF) and non spin flip (NSF), but only one of these two components (the SF) would be expected to contain pinch points (see supporting online text). The previous unpolarised studies~\cite{newprl,meeprb,meenp,clancy,tarasprl} measured the sum of the SF and NSF scattering. 

A single crystal of Ho$_2$Ti$_2$O$_7$ was mounted on the D7 polarization analysis spectrometer at the Institut Laue-Langevin, France, to allow us to map diffuse scattering over the entire $(h,h,l)$ plane. 
Here $x$ ($y$) is defined to be parallel (perpendicular) to the scattering vector ${\bf Q}$ in the horizontal scattering plane. With vertical incident neutron polarization, the SF and NSF cross sections yield information on $S^{yy}({\bf Q})$ and $S^{zz}({\bf Q})$ respectively (see supporting online text ). 

\begin{figure*} [h]
\begin{center}
      \includegraphics[trim=19 180 50 200,clip=true,scale=0.85]{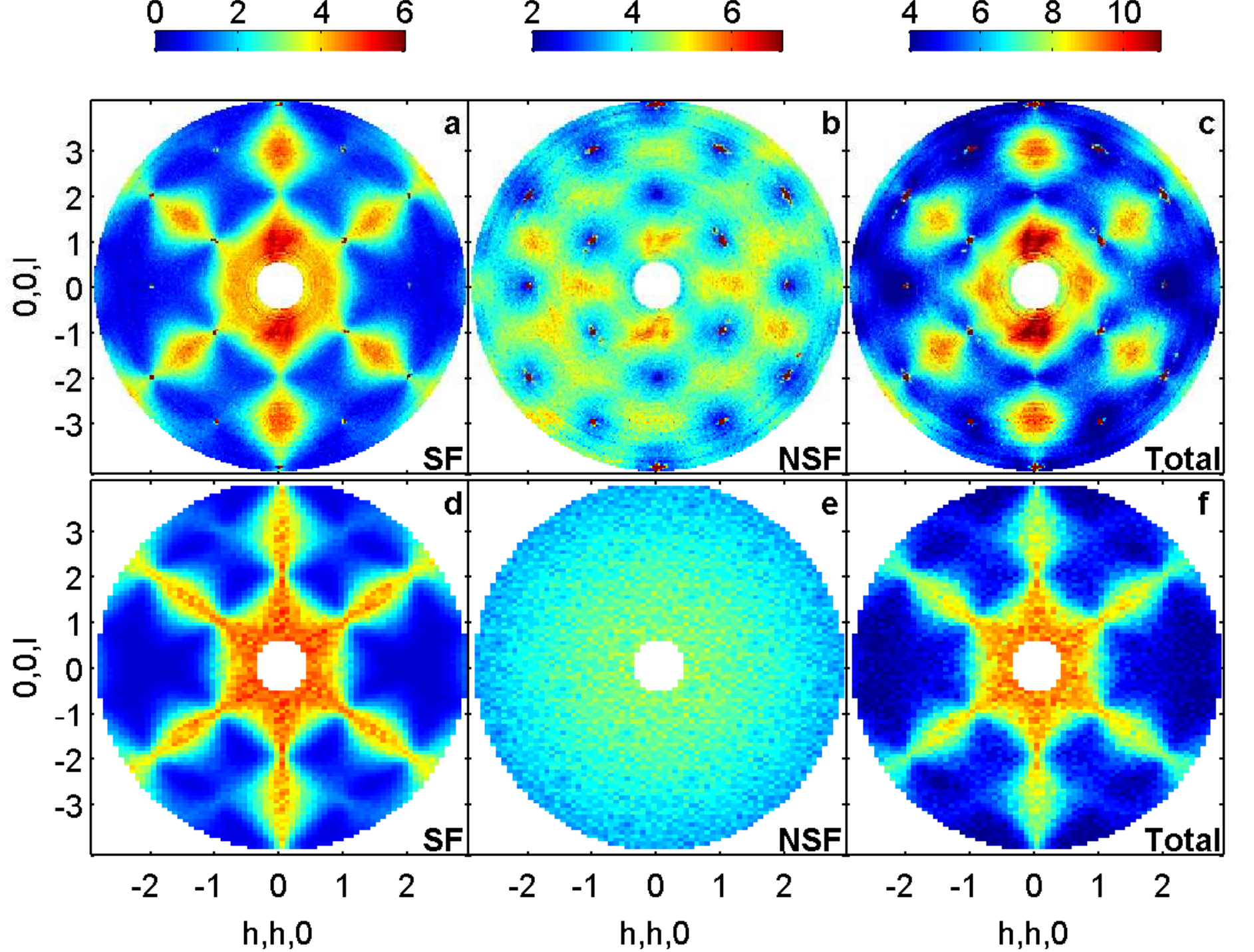}
 \caption{Diffuse scattering maps from spin ice, Ho$_2$Ti$_2$O$_7$. Experiment (top row, a-c) versus theory (bottom row. d-f)  a: Experimental SF scattering at T = 1.7 K with pinch points at (0,0,2), (1,1,1) and (2,2,2) etc. b: the NSF scattering. c: the sum, as would be observed in an unpolarized experiment (eg. Refs~\cite{newprl,clancy}). Sharp spots are nuclear Bragg peaks and  the data have been corrected for absorption.  
 d: The SF scattering obtained from Monte Carlo simulations of the near neighbour model with $12\times12\times12$ supercell, scaled to match the experimental data.  e: The NSF scattering from this model is $\mathbf{Q}$-independent, modulated only by the Ho$^{3+}$ magnetic form factor.  f: The total scattering of the near neighbour spin ice model.}
   \label{maps}
   \end{center}
\end{figure*}

\begin{figure} [h]
\begin{center}
   \includegraphics[trim=150 15 80 5,clip=true,scale=0.6,angle=0]{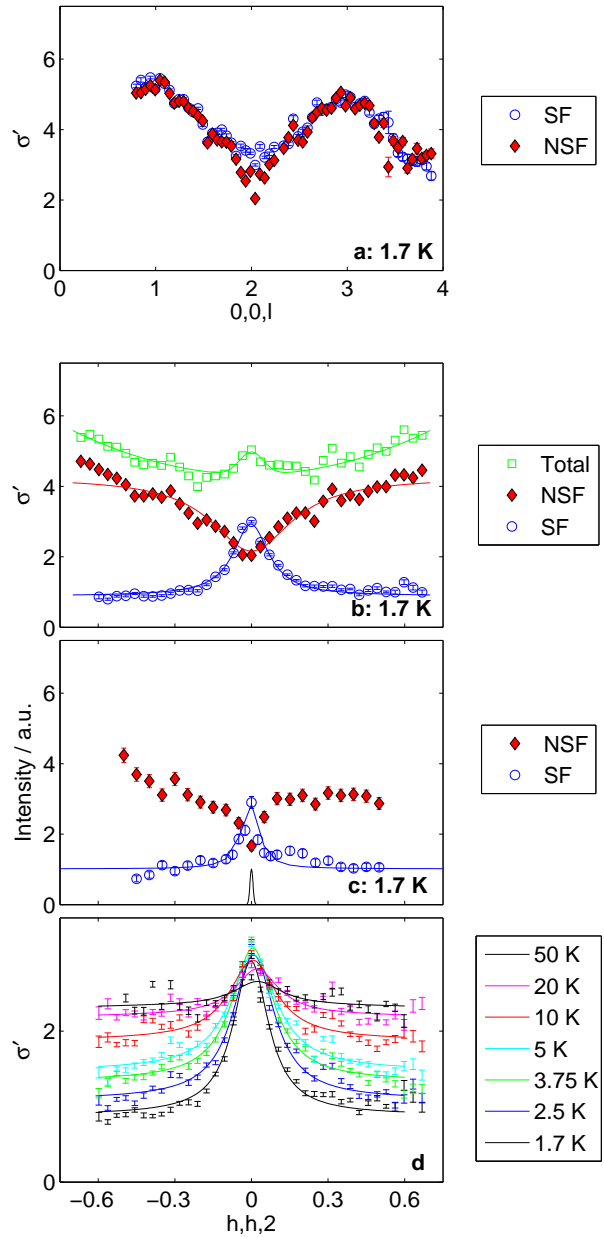} 
    \caption{Line shape of the pinch point. Top (a): radial scan on D7 through the pinch point at (0,0,2) ($\sigma$' is the neutron scattering cross section: see supporting online text for its precise definition). Second (b): the corresponding transverse scan, where the lines are Lorentzian fits. Third (c): higher resolution data from IN12, where the line is a resolution corrected fit to the pinch point form Eqn. (\ref{iceEq}): the resolution width of IN12 is indicated as the central Gaussian.   Bottom (d):  SF scattering at increasing temperatures (the lines are Lorentzians on a background proportional to the Ho$^{3+}$ form factor). }
   \label{pp}
   \end{center}
\end{figure}

In Fig.~\ref{maps} a-c we show the scattering in the SF and NSF channels at $T= 1.7$ K and the total, which would be observed in an unpolarized experiment and can therefore be compared to the observations and calculations of Refs.~\cite{newprl} and ~\cite{clancy}.  Pinch points are clearly visible in the SF cross section at the Brillouin zone centres  (0,0,2), (1,1,1) and (2,2,2), while the NSF scattering does not contain pinch points. The total scattering (SF+NSF) reveals the pinch points only very weakly, if at all (Fig.~\ref{maps}c), as the NSF component dominates near the zone centre.  This is explicitly illustrated in Fig.~\ref{pp}b, where cuts across the zone center show that the strong peak at the pinch point in the spin flip channel is only weakly visible in the total.  The use of polarized neutrons thus extracts the pinch point scattering from the total scattering, and the previous difficulty in resolving the pinch point is clearly explained. 

The projective equivalence of the dipolar and near-neighbour spin ice models~\cite{isakov} suggests that above a temperature scale set by the differences between the two models, the scattering from 
Ho$_2$Ti$_2$O$_7$ should become equivalent to that of the near neighbour model.  $T = 1.7$ K should be sufficient to test this prediction as it is close to the temperature of the peak in the electronic heat capacity that arises from the spin ice correlations (1.9 K)~\cite{newprl}.  Simulations of the near neighbour spin ice model are shown in Fig.~\ref{maps}d-f. Here an effective $J = 1.26$ K is chosen to give the best fit to experiment. Note that it differs from the effective $J_{\rm eff} = 1.8 $ K derived from comparing specific heats~\cite{science}, but there is no reason to expect these effective scales to be identical. Referring to Fig.~\ref{maps}, the experimental SF scattering appears to be very well described by the near neighbour model, while the NSF scattering is not reproduced by the theory. However, we have discovered that $S({\bf Q})^{\rm experiment}/S({\bf Q})^{\rm theory}$ is approximately the same function $f({\bf Q})$ for both channels. Thus, since the theoretical NSF scattering function is approximately constant, we find  $f({\bf Q}) \approx S({\bf Q})^{\rm experiment}_{NSF}$. This function may be described as reaching a maximum at the zone boundary and a finite minimum in the zone centre. Using the above estimate of $f({\bf Q})$, the comparison of the quantity $S({\bf Q})_{SF}^{\rm experiment}/f({\bf Q})$ with $S({\bf Q})_{SF}^{\rm theory}$ is considerably more successful.  Differences are less than 5\% throughout most of the scattering map (see supporting online text and Fig.~S\ref{sfig}).

In Fig.~\ref{pp}a and b we show cuts through the pinch point at (0,0,2), at 1.7 K.  It has the form of a low sharp saddle in the intensity.  In the limit of zero temperature, when the ice rules are perfectly obeyed, theory predicts that the pinch point will be a sharp bow tie, with a singularity at the centre: formally $S^{yy}({\bf q}) \propto q_{\perp}^2/(q_{\parallel}^2+q_\perp^2)$. Here $q_{\perp}, q_{\parallel}$ are components of the scattering vector ${\bf Q}$ measured from a reciprocal lattice point. 
In order to better resolve the lineshape of the pinch point we preformed an analogous polarised neutron experiment on the IN12 triple axis spectrometer, which affords higher resolution than D7:  
to achieve this we used incident neutron wavevector $k_i = 1.5$ \AA$^{-1}$ ($E_i = 4.73$ meV) and $30'$ collimators before and after the sample. To compare with theory we use an approximation to an analytic expression given by Henley~\cite{clh} (see also ref. \cite{yb}). In the vicinity of the (0,0,2) pinch point this becomes 
\begin{equation}\label{iceEq}
S^{yy}(q_h,q_k,q_l) \propto \frac{q_{l-2}^2+\xi_{\rm ice}^{-2}}{q_{l-2}^2+q_h^2+q_k^2+\xi_{\rm ice}^{-2}}.
\end{equation}
Here $\xi_{\rm ice}$ is a correlation length for the ice rules that removes the singularity at the pinch point~\cite{yb}. In Fig.~\ref{pp}c we show the high resolution data from IN12, scaled to those of D7.  By convolving with the triple axis resolution function, the IN12 data can be described by this form, with a correlation length $\xi_{\rm ice} \approx 182\pm65$ \AA, representing a correlation volume of about $25000$ spin tetrahedra.  The correlation length rises to $234\pm67$  \AA~at 1.3 K and has a temperature variation consistent with an essential singularity $\sim\exp(B/T)$, with $B =1.7\pm0.1$ K (see Figure 3c). The physical interpretation of $\xi_{\rm ice}$ is discussed below. 

As described above, the scattering in the NSF channel directly visualises corrections to the ideal spin ice behaviour. It is concentrated around Brillouin zone boundaries, as previously observed in unpolarised cross sections for both Ho$_2$Ti$_2$O$_7$~\cite{clancy} and Dy$_2$Ti$_2$O$_7$~\cite{meeprb}. In the latter case, this enhanced zone boundary scattering was well described by fine tuning further neighbour superexchanges in the dipolar model, though it is not a direct consequence of these small terms~\cite{tarasprl}. The NSF scattering (i.e. $f({\bf Q})$) also shows a pronounced symmetric minimum at each Brillouin zone centre, which, with superior resolution, is revealed to be roughly as sharp as the maximum in the SF scattering. This indicates a partial suppression of magnetization fluctuations on the length scale of the ice rule correlation length, yet one that is insufficient to remove the pseudo-dipolar correlations up to that scale. Projective equivalence in Ho$_2$Ti$_2$O$_7$ is thus proved to be robust to the expected $r^{-5}$ corrections~\cite{isakov} and superexchange~\cite{tarasprl}. The distinct structure of the NSF scattering persists to temperatures as high as 10 K, suggesting a simple and generic correction to the near neighbour model.

\begin{figure} [h]
\begin{center}
%\begin{minipage}
   \includegraphics[trim=50 400 210 80,clip=true,scale=0.7]{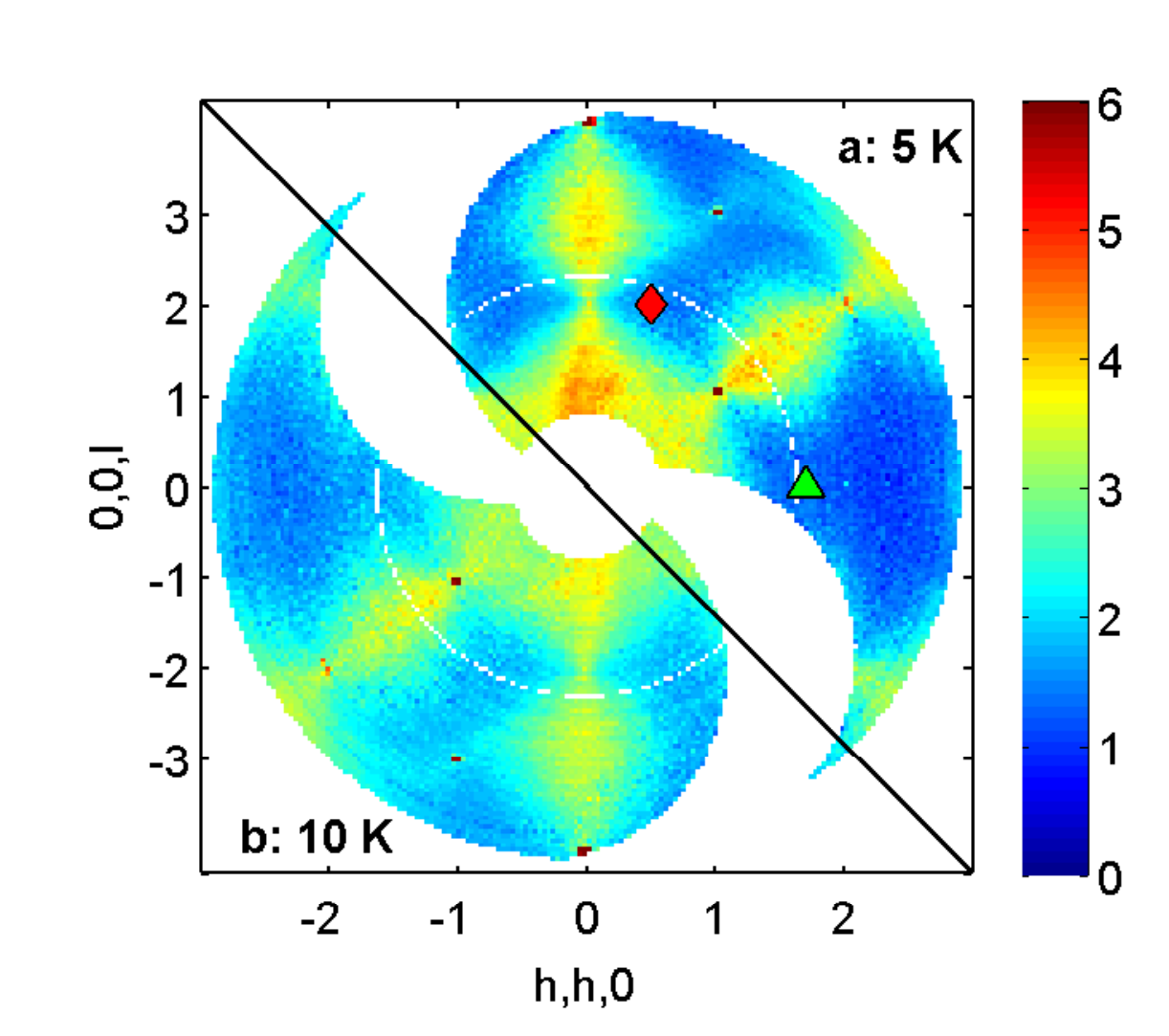}
   \includegraphics[trim=90 220 210 235,clip=true,scale=0.65]{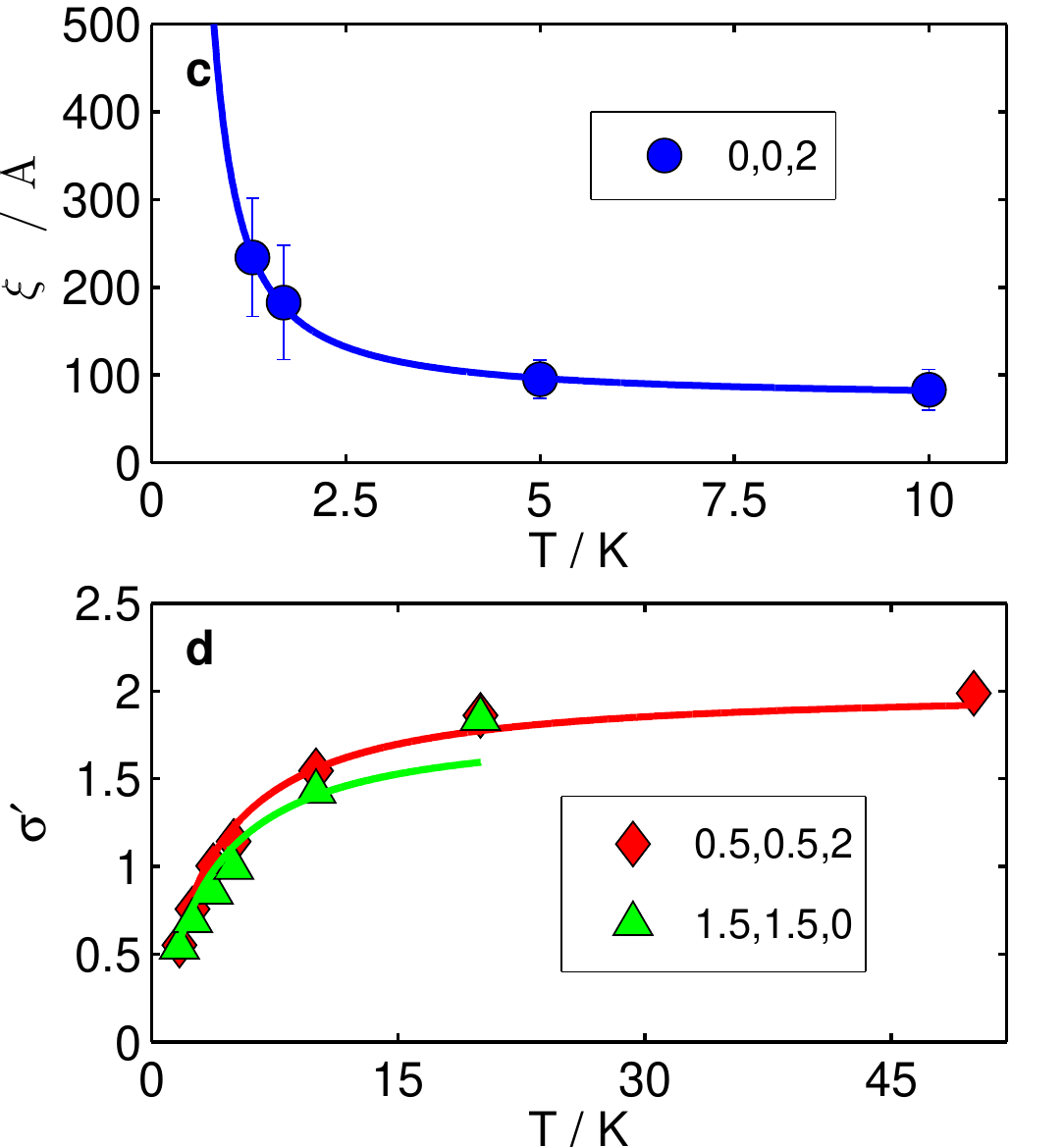}
    \caption{Temperature dependence of defect scattering. a: the SF scattering at 5 K.  b: the spin flip scattering at 10 K.  The rising temperature washes out the sharp pattern of 1.7 K.  In particular, simulations suggest that in the limit of perfect ice rule enforcement, the area around (2,2,0) contains no magnetic scattering.  As the temperature rises, this area is filled in (compare also Fig.~\ref{maps}a). c: the correlation length extracted from the high resolution data is consistent with an exponential divergence of Dirac string length at low temperature.  d: temperature dependence of the scattering at (0.5,0.5,2) (the pinch point background in Fig.~\ref{pp}d) and (1.5,1.5,0) (indicated by corresponding symbols in a).  These parts of the pattern are sensitive to the bound defect population.  The lines are fits to Eqn. \ref{defect} (there is a fitted scale factor), which describes the formation of  single and doubly `charged' ice rule defects (see text).    }
   \label{oneb}
   \end{center}
\end{figure}

In Fig.~\ref{oneb}a we illustrate the general effect of increasing temperature on the SF scattering pattern.  Empty areas of $S^{yy}$ (e.g. near 1.5,1.5,0) are increasingly filled in as the temperature rises, with a  thermal contribution that is independent of wavevector (apart from the Ho$^{3+}$ form factor), indicating 
uncorrelated point defects, or monopoles which remain strongly confined as dipole pairs. This is also illustrated in Fig. \ref{pp}d where, 
with increasing temperature, there is a marked decrease in peak intensity and an increase in the background on which the peak stands.  As shown  in Fig.~\ref{oneb}d, this contribution can be generally fitted by the following form:
\begin{equation}\label{defect}
 I(T) \propto \frac{\exp(-2J/T)+\exp(-8J/T)}{1+\exp(-2J/T)+\exp(-8J/T)},
 \end{equation}
where $I(T)$ is the intensity and $J \equiv J_{\mathrm{eff}} = 1.8$ K, the effective near neighbour exchange appropriate for Ho$_2$Ti$_2$O$_7$~\cite{science}.  The two terms are the cost of creating singly `charged' (monopole) or doubly `charged' thermal defects in the ice rules respectively. 

Experimental evidence for the additional contribution of {\it deconfined} defects is provided by the broadened pinch point scattering, discussed above.   We interpret this length scale as the maximum length of the effective Dirac strings that connect  deconfined magnetic monopoles~\cite{monopoles,ludovic}.  A Dirac string is a line of overturned dipoles connecting two monopoles, i.e. a ferromagnetic fluctuation in spin ice.  At high temperatures the proliferation of bound defects will both disrupt existing strings, and reduce the mean free path for diffusing monopoles, reducing the maximum length in the Dirac string network.  As the temperature is reduced, the thermal defect population decreases and we see the correlation length of the Dirac string network diverges as approximately $\exp(B/T)$, as shown in Fig.~\ref{oneb}c, with the observed value of $B$ very close to the effective exchange $J_{\rm eff} = 1.8$ K~\cite{science}. Such a temperature variation of the correlation length is the same as that of the one dimensional Ising ferromagnet, where the interpretation of the correlation length as an effective maximum string length is exact.  

The observed temperature dependent contributions to the diffuse scattering are thus well described by the sum of two terms, corresponding to confined and deconfined monopoles respectively. The integrated intensity shows that the confined monopoles are by far the dominant contribution at finite temperature, but the importance of the deconfined pairs increases as the temperature decreases, and at low temperatures they are deconfined over macroscopic (at least micron) length scales.  This behaviour is fully consistent with a picture of a weakly dissociated electrolyte where the energy cost of dissociation (including screening) is small compared to the energy of ion pair formation. In this regard, the spin ice thermal defects are fully equivalent to ionic defects in water ice (but note that there is no equivalent of the `Bjerrum defect'~\cite{tchernyshyov}).

In conclusion, we have studied the spin ice Ho$_2$Ti$_2$O$_7$ using polarized neutron scattering.   
We have established the validity of projective equivalence and have quantified the corrections to it. These corrections are summarised by the function $f({\bf Q})$ that dips sharply at the zone centre, raising the question of whether $f({\bf Q})$ can be captured in a generic improvement to near neighbour spin ice that still obviates a full treatment of the microscopic Hamiltonian. 
More importantly, we have established that Ho$_2$Ti$_2$O$_7$ exhibits an almost ideal magnetic Coulomb phase, the quasiparticle vacuum for Castelnovo {\it et al.}'s magnetic monopoles~\cite{monopoles,ludovic}.  We have estimated the length of the longest Dirac strings to rise to macroscopic scales as the temperature passes below 1 K.

%\section{}
%\subsection{}
\begin{scilastnote}
%\section{Acknowledgements}
\item We thank P. Holdsworth and L. Jaubert (ENS-Lyon), M. Harris and J. T. Chalker (Oxford), A. Fisher (LCN), L.P. Regnault (ILL) and R. Moessner (Dresden) for valuable discussions, and L. Fodinger, J. Previtali, O. Losserand and A. Filhol of the ILL for technical assistance with D7, IN12, cryogenics and computation respectively.  We acknowledge the EPSRC and STFC (UK) for funding.
\end{scilastnote}

\bibliography{fennell_arx}{}
\bibliographystyle{science}
\newpage

\newpage

\newpage

\newpage

\clearpage

{\bf \large Materials and Methods} 

{\em Crystal growth:} polycrystalline Ho$_2$Ti$_2$O$_7$ (space group Fd$\bar{3}$m) was synthesized from high purity Ho$_2$O$_3$ and TiO$_2$ powders by reacting in air at 1250 $^\mathrm{\circ}$C for 3 days with intermediate
grindings.  A crystal boule was grown using a four mirror optical floating
zone furnace with a growth speed of 1-2mm/h and a rotation speed of 25
rpm under flowing oxygen gas. The single crystal used was cut from this boule and its quality checked by x-ray diffraction.  The sample is approximately cylindrical with radius $\approx 3$ mm and length $\approx 45$ mm.

{\em Neutron Scattering experiments on D7:} The crystal was oriented with the $[1\bar{1}0]$ axis vertical (with the boule tilted at some 20$^\mathrm{\circ}$ away from vertical) and clamped to a copper holder.  In order to minimize the incoherent scattering background, the use of glue was avoided.  The temperature was controlled using a standard Orange cryostat.  Details of the polarization analysis system of D7 can be found in Ref.~S1.

With $z = [1\bar{1}0]$ vertical, the $(h,h,l)$ wavevectors perpendicular to this formed a conventional $xy$ scattering plane. The incident neutron energy was 3.5 meV and the incident neutron polarization $\mathbf{\hat{P}}$ was vertical. As in previous studies~\cite{newprl,meeprb,meenp,clancy,tarasprl}, the static approximation (i.e. an appropriate integration over energy) was ideally obeyed, so the scattering function measured the arrangement of spins at an instant in time, averaged over an ensemble of identical systems. Defining $x$ as parallel to ${\bf Q}$ and $y$ orthogonal to $x$ and $z$, the SF and NSF cross sections are proportional to $S^{yy}({\bf Q})$ and $S^{zz}({\bf Q})$ respectively. The tetrahedral basis of the pyrochlore stucture has four spins: for a general scattering vector all four spins contribute to $S^{yy}({\bf Q})$, but only two contribute to $S^{zz}({\bf Q})$ as the  Ising anisotropy means that the other two have no component on $z$. This has the strong consequence that pinch point singularities are expected in $S^{yy}({\bf Q})$ but not in $S^{zz}({\bf Q})$, as only $S^{yy}$ will contain the full set of ice rule correlations. The cross section for unpolarised neutron scattering is proportional to $S^{yy}({\bf Q}) + S^{zz}({\bf Q})$. Other contributions to the scattering (nuclear Bragg,  nuclear thermal diffuse,  nuclear spin and isotopic incoherent scattering) have a negligible effect on the results presented here, so will not be discussed further.  

We measured the non-spin flip and spin flip cross sections with the polarization parallel to $z$ by rotating the crystal ($\Omega$).  The data in Fig.~\ref{maps} of the paper were obtained by 0.5$^\mathrm{\circ}$ step rotations and with two positions ($\theta$) of the multi-detector banks to eliminate gaps in the scattering map.  
At higher temperatures, only one detector bank position was measured, resulting in the gap that can be seen in Fig.~\ref{oneb}a-b.
The statistics around the (0,0,2) pinch point were improved by a factor of 2 by extended counting at all temperatures.  The classical $\parallel-\perp$ polarization analysis technique can be applied at a single value of $|Q|$ enabling a complete separation of scattering contributions in anisotropic materials, not generally possible in the multidetector $xyz$ measurement.  We measured this cross section at a single point at 20 K to confirm our normalization procedure.

The data were corrected as described in Ref. S1.  In brief,  the instrumental background was subtracted, an amorphous silica standard was used  to normalize the supermirror analyzer efficiencies and a known quantity of vanadium to normalize detector efficiencies and provide an absolute scale.  All scattering data from D7 shown in the paper is therefore a measurement of the absolute partial differential cross section  $\sigma' \equiv d\sigma/d\Omega$ in barn steradian$^{-1}$ (formula unit)$^{-1}$.    Additionally the data were corrected for absorption.  Path dependent attenuation coefficients were calculated for every point in the $\Omega-\theta$ map using the $UB$-matrix approach outlined in Refs. S2 and S3.  After this procedure the $\parallel-\perp$ measurement confirmed that the isotopic incoherent scattering was as expected for Ho$_2$Ti$_2$O$_7$.  The spin incoherent scattering was somewhat larger than calculated, but only a {\em very} small amount of hydrogenous material would be required to produce this effect.  In Fig. 1a sharp intensity is visible at the centre of the pinch points at (1,1,1), (2,2,2) and (0,0,4).  This is due to finite polarization analyzer efficiency and is noticable only at the position of a very intense nuclear Bragg peak.  As the temperature is lowered the sample slightly depolarizes the beam, but we confirmed that the total intensity of the nuclear peaks summed over both channels is constant, and corrected for the depolarization effect by comparing the flipping ratio of the Bragg peak with the ideal value for the quartz standard at the same position.

{\em Neutron scattering experiments on IN12:} The crystal was mounted as described above, but in a dilution refrigerator insert (we do not discuss the low temperature data as depolarization effects become more severe).  IN12 was configured with a focussing graphite monochromater, supermirror bender, Heusler analyzer and open-30-30-open collimators.  The convolution of equation 1 was achieved by using a raytracing Monte Carlo integration algorithm described in Ref S4.

{\em Monte Carlo simulations:} We simulated the near neighbour spin ice model using a Metropolis Monte Carlo algorithm combining single spin and loop dynamics.  The single spin flip dynamics establish equilibrium at high temperature and the loop moves maintain ergodicity at low temperature, as described in Ref. S5 and references therein.  For comparison with neutron scattering we averaged two simulations of $12\times 12\times 12$ pyrochlore unit cells, each contributing 200 spin configurations.  We estimated the correlation time and ensured that spin configurations collected from each simulation were separated by at least this many Monte Carlo sweeps.  The calculated polarized cross sections are multiplied by the form factor of Ho$^{3+}$ (Ref. S6).  

To compare directly with the experimental data, we subtract the estimated spin incoherent cross section from the data and scale the simulated result.  When invoking $f(Q)$, we divide the experimental spin flip scattering by the non-spin flip and then scale the simulated data.  In this case it is not necessary to include the magnetic form factor as it is divided out of the experimental data.  We found no improvement in this process by modifying the scale or baseline of the non-spin flip data and this is unsurprising since Fig~\ref{pp}a of the paper shows them to be essentially identical in certain directions, except at the zone centre.  The comparison requires the exclusion of nuclear Bragg peaks and the zone centres, where the departure of the non-spin flip from the spin flip creates an artificial peak in the divided data.  We quantify the goodness of the fit using a difference map, showing here the results for the best fitting simulation.  It is clear that the match is good throughout the scattering map, except for a specific collection of wavevectors, generally transverse to a pinch point, where the approximation of a single $f(\mathbf{Q})$ must be poor.  

 \noindent S1. J. R. Stewart, {\it et al., J. Appl. Cryst.} {\bf 42}, 69 (2009)\\
 S2. W. R. Busing, H. L. Levy, {\it Acta Cryst.} {\bf 10}, 180 (1957)\\
 S3. W. R. Busing, H. L. Levy, {\it Acta Cryst.} {\bf 22}, 457 (1967)\\
 S4. J. $\mathrm{\check{S}}$aroun, J. Kulda, {\it Physica B} {\bf 234-236}, 1102 (1997)\\
 S5. R. G. Melko and M. J. P. Gingras, {\it J. Phys.: Condens. Matter} {\bf 16}, R1277 (2004).\\
 S6. P. J. Brown, {\it Magnetic Form Factors,}, vol. C of {\it International Tables for Crystallography} (Kluwer Academic, Dodrecht, 1992).

 \newpage
 
 \renewcommand{\thefigure}{\arabic{figure}}
 \addtocounter{figure}{-3}
\begin{figure}
\renewcommand{\figurename}{S}
%\centering
\begin{minipage}{0.5\textwidth}
   \includegraphics[trim=80 240 90 20,clip=true,scale=0.55]{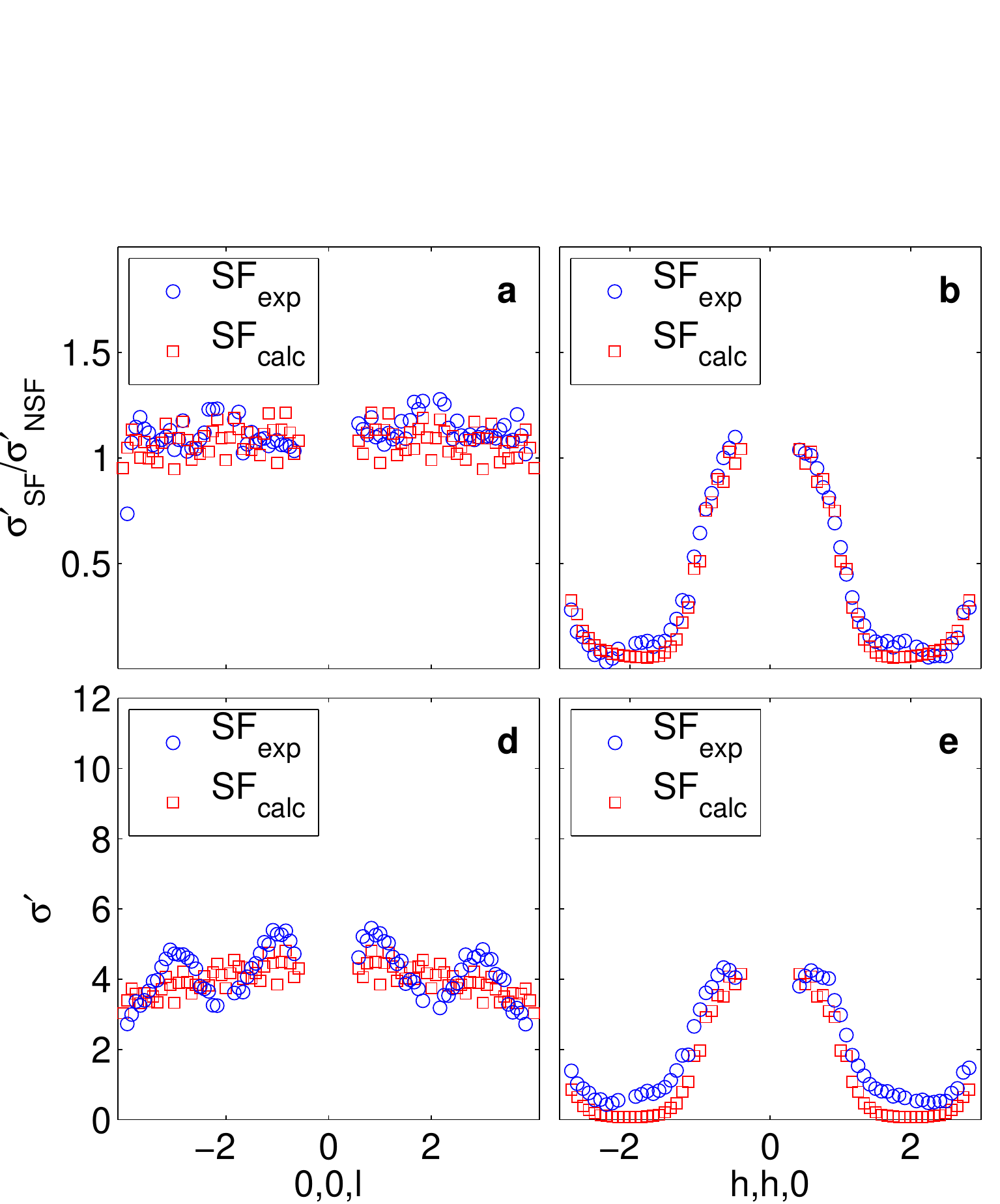}
   \end{minipage}%
   \begin{minipage}{0.5\textwidth}
   \includegraphics[trim=320 200 40 150,clip=true,scale=0.8]{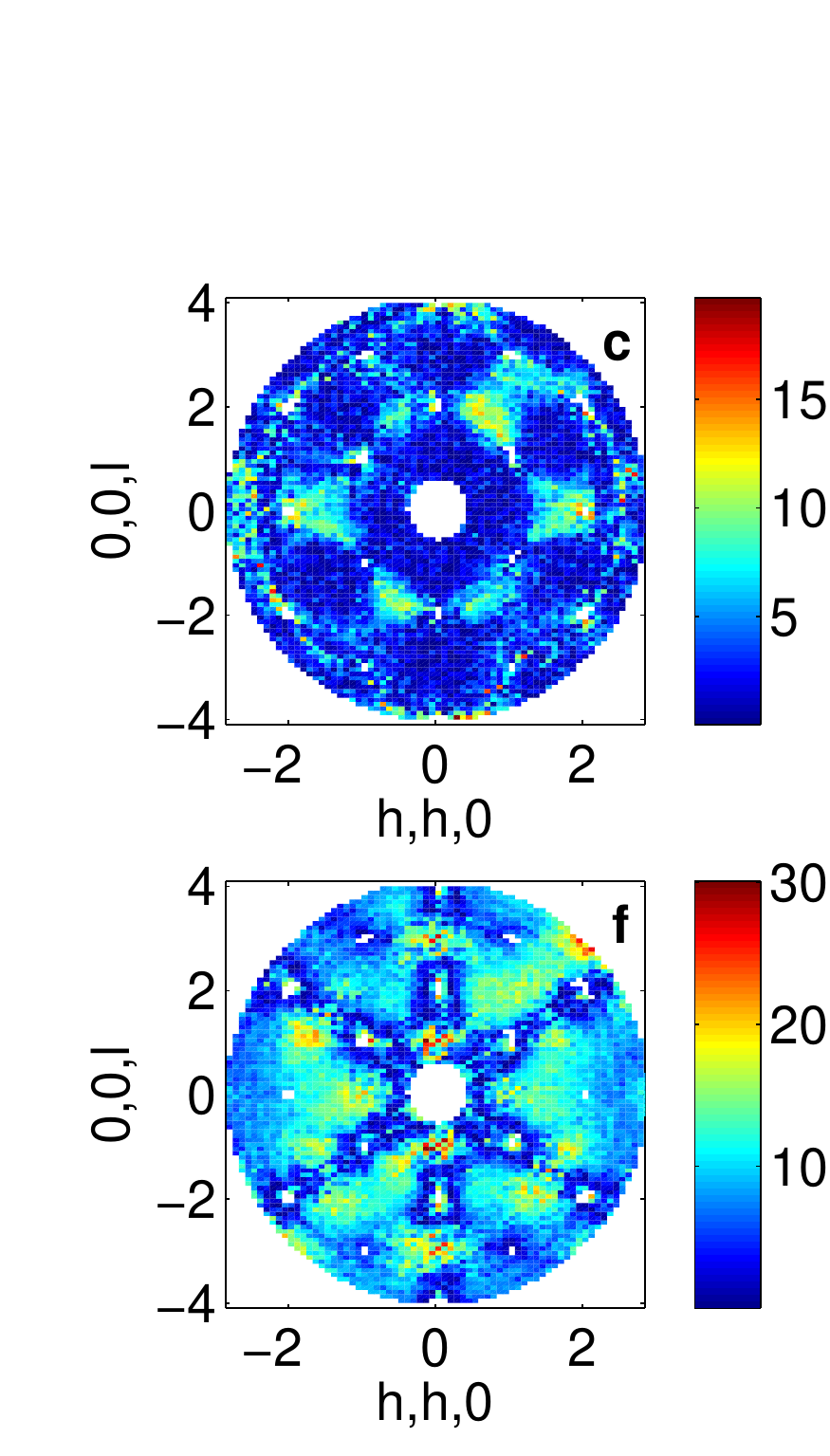}
   \end{minipage}
    \caption{Comparison of the experimental data and Monte Carlo simulations.  In the top panels the comparison includes $f(Q)$, i.e the experimental spin flip scattering is divided by the experimental non-spin flip scattering, and the simulation scaled.  a and b show cuts along $(0,0,l)$ and $(h,h,0)$ respectively. In c we show the percentage difference between the calculated and experimental scattering.  In the bottom panels we show the direct comparison (as in Fig. 1 of the paper) in which the simulation is simply scaled to the experimental data.  Because the experimental data is modulated along $(0,0,l)$, but the simulation is not, significant mismatch appears.  }
  \label{sfig}
%   \end{center}
\end{figure}

\end{document}